\documentclass{amsart}
\usepackage{amsmath}                                %
\usepackage{mathrsfs}
\usepackage{amssymb}
\usepackage[dvips]{graphicx}
\begin{document}

\begin{center}
{\Large\bf Topology Change of Black Holes}\\
\bigskip
{\large\sc Daisuke Ida${}^1$ and Masaru Siino${}^2$
}\\
\medskip
$^1$Department of Physics, Gakushuin University, Tokyo 171-8588, Japan.\\
$^2$Department of Physics, Tokyo Institute of Technology, Tokyo 152-8550, Japan.\\

\vspace{5pt}
(April 1, 2007)\\
\vspace{5pt}
{Abstract.}
\end{center}
The topological structure of the event horizon has been 
investigated
in terms of the Morse theory.
The elementary process of 
topological evolution
 can be understood as a handle attachment.
It has been found that there are certain constraints on 
the nature of black hole topological evolution:
(i) There are $n$ kinds of handle attachments
in $(n+1)$-dimensional black hole space-times.
(ii) Handles are further classified as either of black or white type,
and  only black handles appear in real black hole space-times.
(iii) The spatial section of an exterior of the black hole region is always connected.
As a corollary, it is shown that the formation of a black hole
with an $S^{n-2}\times S^1$ horizon 
from that with an $S^{n-1}$ horizon must be non-axisymmetric in asymptotically flat
space-times.

\bigskip

\section{Introduction}

Black holes in space-times of greater than or equal to five dimensions
 have rich topological structure.
According to the well-known results of Hawking concerning the topology of black holes 
in four-dimensional space-time,
the apparent horizon or the spatial section of the 
stationary event horizon is necessarily diffeomorphic
to a 2-sphere.~\cite{Hawking:1971vc,HE}
This follows from the fact that the total curvature, which is the integral of the
intrinsic scalar curvature over the horizon, is positive under 
the dominant energy condition and from the Gauss-Bonnet theorem.
Alternative and improved proofs of Hawking's theorem have been given by several
authors.~\cite{Ga74,Chrusciel:1994tr,Jacobson:1994hs,Browdy:1995qu}
However in higher dimensional space-times, an apparent horizon or the spatial section of the 
stationary event horizon may not be
a topological sphere,~\cite{Cai:2001su,Helfgott:2005jn,Galloway:2005mf,Galloway:2006ws}
because 
the Gauss-Bonnet theorem does not hold in such cases.
Nevertheless, the positivity of the total curvature
of the horizon still holds.
This puts certain topological restrictions on the black hole topology, 
though they are rather weak.
For example, the apparent horizon in five-dimensional space-time can consist of
finitely many connected
sums of  copies of $S^3/\Gamma$ and  copies of $S^2\times S^1$.
In fact,  exact solutions representing a black hole space-time possessing a horizon
of nonspherical topology have recently been found in five-dimensional general relativity.
When such black holes with nontrivial topologies are regarded as being formed in the course of
gravitational collapse, 
questions regarding the evolution of the topology of black holes naturally arise.
Our purpose here is to understand the time evolution of the topology of event horizons
 in a general setting.
The relation between the crease set, where the event horizon
is nondifferentiable, and the topology of the event horizon is
studied in Refs.~\cite{Siino:1997ix,Siino:2004xe,Shapiro:1995rr} 
for four-dimensional space-times.
In the present work,
we carry out a systematic investigation and 
find useful rules to determine  admissible processes of topological evolution
for time slicing of a black hole.

Our approach is to utilize the Morse theory~\cite{Mil65,Ta78} in differential topology.
The Morse theory is useful for the purpose of understanding the topology of smooth manifolds.
The basic tool used in this approach is a smooth function on a
differentiable manifold.
The event horizon, however, is not a differentiable manifold but has a wedge-like structure
at the past endpoints of the null geodesic generators of the horizon.
For this reason, we first smooth the wedge. Then, the smooth time function 
which is assumed to exist
plays the role of the Morse function on the smoothed event horizon.
According to the Morse theory, the topological evolution of the event horizon can then be 
decomposed into elementary processes called ``handle attachments.''
In such a process, starting with a spherical horizon, one adds several handles,
 each characterized by the
index of the critical points of the Morse function, 
which is an integer ranging from $0$ to $n$ 
(the dimension of the smoothed horizon as a differentiable manifold).

The purpose of the present article is to show that 
there are several constraints on the handle attachments 
for real black hole space-times.

\section{The Morse theory for event horizons}\label{s2}
Let $M$ be an $(n+1)$-dimensional asymptotically flat space-time.
We require the existence of a global time function $t: M\to {\mathbb R}$
that is smooth and 
has an everywhere time-like and future-pointing gradient.
The event horizon $H$ is defined as the boundary of the causal past of the future null
infinity $H=\partial J^-({\mathscr I}^+)$.~\cite{HE}
We treat the event horizon defined with respect to a single asymptotic end,
unless otherwise stated. 
In other words, the future null infinity, ${\mathscr I}^+$, is assumed to be connected.
The black hole region ${\mathscr B}$ is defined as the interior region of $H$, 
specifically, as ${\mathscr B}=M\setminus J^-({\mathscr I}^+)$,
 and the exterior region ${\mathscr E}$ of
the black hole region is its complement, ${\mathscr E}={\rm int}(J^-({\mathscr I}^+))$.
We refer to the intersection of the black hole region
and the time slice $\Sigma(t_0)=\{t=t_0\}$
as the black hole ${\mathscr B}(t_0)={\mathscr B}\cap \Sigma(t_0)$
at time $t=t_0$. The exterior region at time $t=t_0$ is, accordingly, written
${\mathscr E}(t_0)={\mathscr E}\cap \Sigma(t_0)$.

One of most basic properties of the event horizon is that 
it is generated by null geodesics without future endpoints. 
In general, the event horizon is not smoothly imbedded into the space-time manifold $M$, 
but it has a wedge-like structure at the past endpoints of the null geodesic generators, 
where distinct null geodesic generators intersect.
We call the set of past endpoints of null geodesic generators of $H$,
from which two or more null geodesic generators emanate, 
the crease set $S$.~\cite{Siino:1997ix,Siino:2004xe}
When no crease set $S$ exists between the time slices $t=t_1$ and $t=t_2$, the
null geodesic generators of $H$ naturally define a diffeomorphism
$\partial{\mathscr B}(t_1)\approx \partial{\mathscr B}(t_2)$.
Hence, the topological evolution of a black hole can take place only when the time slice
intersects the crease set $S$.
Of course, the event horizon itself is a gauge-independent object.
Nevertheless, we often understand the dynamics of space-time by scanning it 
along time slices. 
Thus, the topological evolution of a black hole depends on the time function.

\begin{figure}[t]
\includegraphics[width=.5\linewidth]{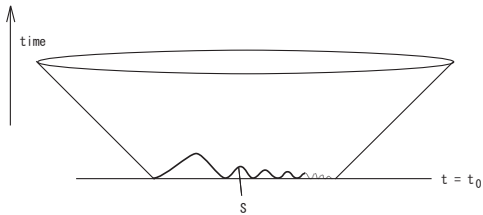}
\caption{An example in which no smoothing procedure makes
$t|_{\widetilde {H}}$ a Morse function on $\widetilde H$.
Here, the intersection of the crease set $S$ of the event horizon and $t=t_0$
hypersurface has an accumulation point.}
\label{accum}
\end{figure}

It is expected that Morse theory~\cite{Mil65} provides useful techniques 
to analyze such a process of topological evolution.
Because the Morse theory 
is concerned with  functions on smooth manifolds,
we first regularize $H$ around the crease set $S$.
The event horizon is not necessarily smooth, even on $H\setminus S$, in the case that the
future null infinity ${\mathscr I}^+$ has a pathological structure.~\cite{Chrusciel:1996tw}
Here it is assumed that $H$ is smooth on $H\setminus S$.
Then, small deformations of $H$ near the crease set $S$ will make $H$ a smooth hypersurface
$\widetilde H$ in $M$, 
while ${\widetilde {\mathscr B}}(t_0)$ remains deformed in such a manner that
$\partial\widetilde{{\mathscr B}}(t_0)=\widetilde H\cup \Sigma(t_0)$ holds and
${\widetilde {\mathscr B}}(t_0)$ remains homeomorphic to the original black hole
for all $t_0\in {\mathbb R}$.
This deformation is assumed to be  such that the
time function $t|_{\widetilde{H}}$, which is the restriction of $t$ on $\widetilde H$,
gives a Morse function on $\widetilde H$ that has only
nondegenerate critical points, where the gradient of $t|_{\widetilde{H}}$ 
defined on $\widetilde H$ becomes zero and where
also the Hessian matrix $(\partial_i\partial_j t|_{\widetilde H})$ 
of $t|_{\widetilde{H}}$ 
is nondegenerate.
Though this assumption should hold for a wide class of systems, it does not always hold.
Figure~\ref{accum} gives an example for which no smoothing procedure makes the induced
time function $t|_{\widetilde {H}}$ a Morse function on $\widetilde H$,
because the intersection of the crease set $S$ of the event horizon and the
$t=t_0$ hypersurface has an accumulation point.
It is highly nontrivial to determine whether such a smoothing procedure is generically possible.
It is, however, not easy nor the primary purpose of this article
to assertain the realm of validity of the assumption, 
and therefore we make this assumption 
without inquiring into its validity.

According to the Morse Lemma, there is a local coordinate system $\{x^1,\cdots,x^n\}$ 
on $\widetilde H$ in the neighborhood of the critical point $p\in \widetilde{H}$ such that
the restriction $t|_{\widetilde{H}}$ of the time function $t$ on $\widetilde{H}$ 
takes the form
\begin{eqnarray*}
t|_{\widetilde{H}}(x^1,\cdots,x^n)=t(p)-(x^1)^2-\cdots-(x^\lambda)^2+(x^{\lambda+1})^2+\cdots+(x^n)^2.
\end{eqnarray*}
The integer $\lambda$, ranging from $0$ to $n$, is called the index of the critical point $p$.
The topology of the black hole $\widetilde{{\mathscr B}}(t)$ changes when $\Sigma(t)$ pass through  critical points,
or equivalently, when the time function $t$ takes  critical values.
This implies that  critical points appears only near the crease set $S$.

\begin{figure}[t]
\includegraphics[width=0.8\linewidth]{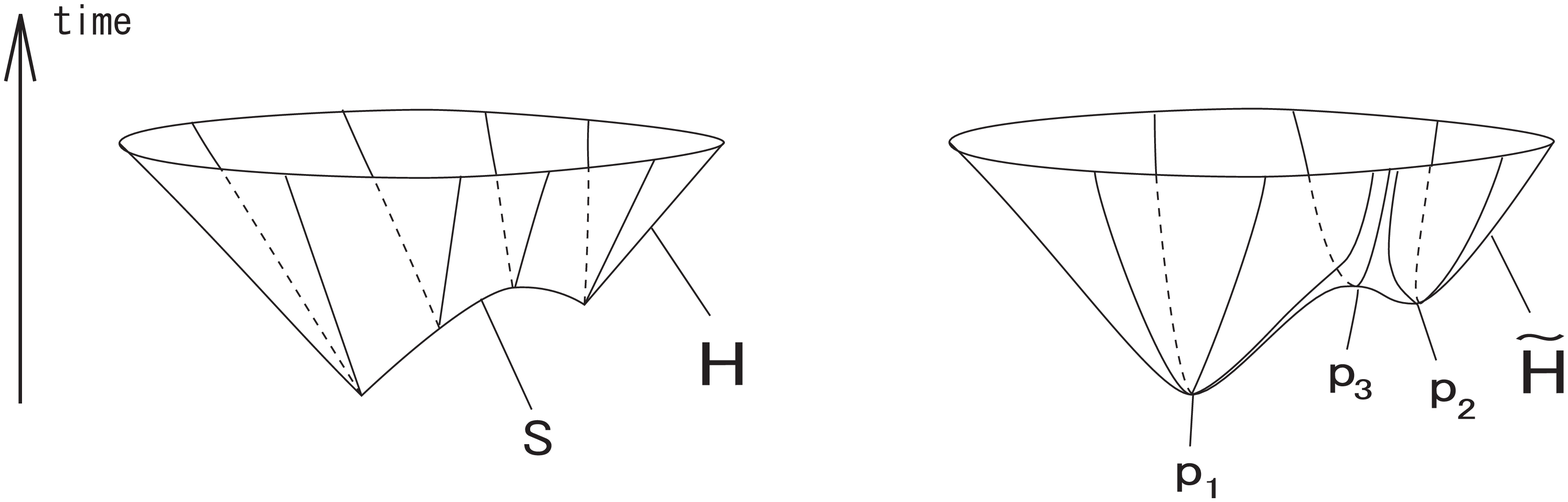}
\caption{The smoothing procedure of the event horizon $H$.
The gradient-like vector field on $\widetilde H$ can be constructed through a slight
deformation of the null geodesic generators of $H$.
Here, the effect of the crease set $S$ has been replaced by 
that of the critical points $p_1$, $p_2$ and $p_3$.}
\label{isolation}
\end{figure}

\begin{figure}[t]
\includegraphics[width=0.7\linewidth]{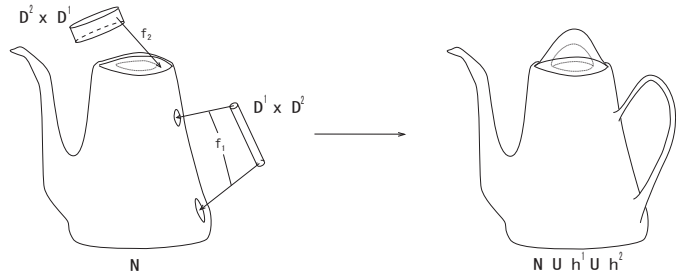}
\caption{
The attachment of a $1$-handle and a $2$-handle to a $3$-manifold $N$
creates a new  $3$-manifold
$N\cup h^1\cup h^2$.
}\label{pot}
\end{figure}

The gradient-like vector field for $t|_{\widetilde{H}}$ is defined to be the tangent 
vector field $X$ on $\widetilde H$
such that $Xt|_{\widetilde{H}}>0$ holds on $\widetilde H$,
except for critical points, and has the form
\begin{eqnarray*}
X=-2x^1{\partial\over\partial x^1}-\cdots-2x^\lambda{\partial\over\partial x^\lambda}
+2x^{\lambda+1}{\partial\over\partial x^{\lambda+1}}+\cdots
+2x^{n}{\partial\over\partial x^{n}}
\end{eqnarray*}
near the critical point of index $\lambda$, in terms of the standard coordinate system
appearing in the Morse Lemma. We choose a gradient-like vector field $X$ such that
it coincides with the future-directed tangent vector field of null geodesic generators 
of $H$, except in a small neighborhood of the crease set $S$~(Fig.~\ref{isolation}).
\begin{figure}[t]
\includegraphics[width=0.7\linewidth]{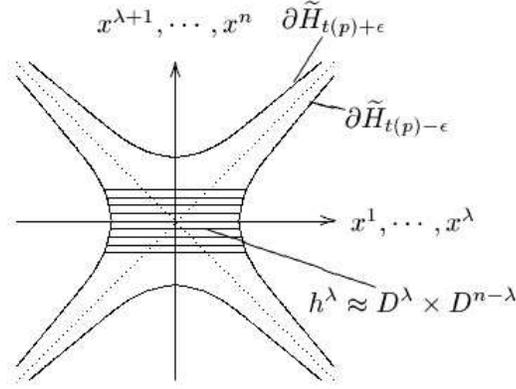}
\caption{The local structure around the critical point $p$ of index $\lambda$.
  It can be seen that $\widetilde H_{t(p)+\epsilon}$ is homeomorphic to
$\widetilde H_{t(p)-\epsilon}$ with a $\lambda$-handle attached.
}\label{critical}
\end{figure}
The effect of a critical point $p$ of index $\lambda$ is equivalent to the
attachment of a $\lambda$-handle.~\cite{Mil65,Ta78}
The handlebody is just a topological $n$-disk $D^n\approx I^n$ $(I=[0,1])$,  
but it is regarded as the product space
$D^n\approx D^\lambda\times D^{n-\lambda}$ (Fig.~\ref{critical}).
The $\lambda$-handle attachment to an $n$-dimensional manifold $N$ with a boundary
consists of the set $h^\lambda=(D^\lambda\times D^{n-\lambda},f)$, where 
the attaching map $f$ induces the imbedding of 
$\partial D^\lambda \times D^{n-\lambda}\subset \partial D^n$
into $\partial N$~(Fig.~\ref{pot}). The new manifold obtained through 
the $\lambda$-handle attachment to $N$ is given by
\begin{eqnarray*}
N\cup h^\lambda=N\cup (D^\lambda\times D^{n-\lambda})/(x\sim f(x)),
~~~(x\in \partial D^\lambda\times D^{n-\lambda}).
\end{eqnarray*}
Let us denote by $\widetilde H_{t_0}$ the $t\le t_0$ part of $\widetilde H$.
Then, $\widetilde H_{t(p)+\epsilon}$ $(\epsilon>0)$ just above
the critical point $p$ of index $\lambda$
is homeomorphic (in fact diffeomorphic, taking account of the smoothing procedure)
to that just below $p$, $\widetilde H_{t(p)-\epsilon}$ attached with a $\lambda$-handle,
\begin{eqnarray*}
\widetilde H_{t(p)+\epsilon}\approx \widetilde H_{t(p)-\epsilon}\cup h^\lambda,
\end{eqnarray*}
if there are no other critical points satisfying $t(p)-\epsilon\le t\le t(p)+\epsilon$.
The handlebody itself is denoted by $h^\lambda$ as well.

\begin{figure}[t]
\includegraphics[width=0.3\linewidth]{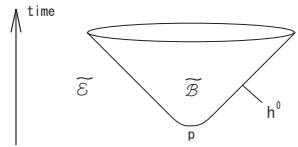}
\caption{The emergence of a black hole through a $0$-handle attachment.}\label{b0-handle}
\end{figure}
\begin{figure}[t]
\includegraphics[width=0.3\linewidth]{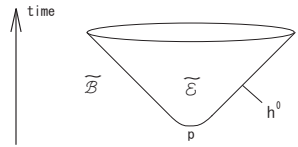}
\caption{The emergence of a bubble in the black hole region by $0$-handle attachment,
which does not occur in the real black hole space-times.}\label{w0-handle}
\end{figure}

Let us consider several examples.
The $0$-handle attachment does not need an attaching map $f$.
It simply corresponds to the emergence of the $(n-1)$-sphere $S^{n-1}\approx\partial D^n$
 as a black hole horizon 
$\partial{\mathscr B}(t)$.
A typical example is the creation of a black hole (Fig.~\ref{b0-handle}):
A black hole always emerges as $0$-handle attachment. 
The other possiblity is the creation of a bubble that is subset of $J^{-}({\mathscr I}^+)$ 
in a black hole region (Fig.~\ref{w0-handle}).
One might think that this corresponds to wormhole creation
between the internal and external regions of the event horizon.
Although in the framework of the standard Morse theory on $\widetilde H$, 
these two examples are indistinguishable,  
we below see that the latter process is in fact impossible.

\begin{figure}[t]
\includegraphics[width=0.4\linewidth]{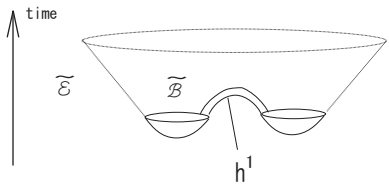}
\caption{The collision of a pair of black holes, creating a single black hole, 
is realized through
$1$-handle attachment.}\label{b1-handle}
\end{figure}

Next, we consider $1$-handle attachment.
A typical example is the collision of two black holes. 
A $1$-handle serves as a bridge connecting black holes, or it corresponds to 
taking the connected sum of each
component of multiple black holes (Fig.~\ref{b1-handle}).

\begin{figure}[t]
\includegraphics[width=0.4\linewidth]{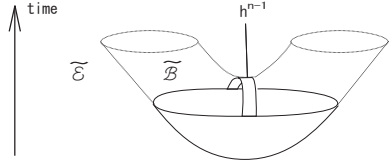}
\caption{The bifurcation of one black hole into two is represeted 
by an $(n-1)$-handle attachment.
This, however, never occurs in  real black hole space-times.}\label{wn-1-handle}
\end{figure}
\begin{figure}[t]
\includegraphics[width=0.5\linewidth]{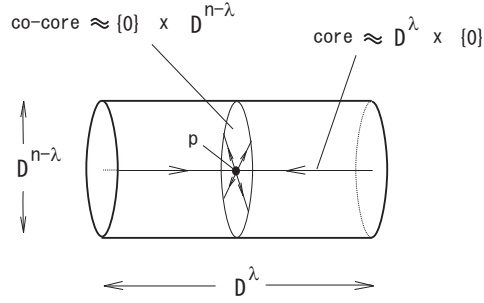}
\caption{The structure of $\lambda$-handle. The core $D^\lambda\times \{0\}$ corresponds
  to the stable submanifold with respect to the flow 
generated by the gradient-like vector field,
and the co-core $\{0\}\times D^{n-\lambda}$ corresponds
to the unstable submanifold.}
\end{figure}

The time reversal of the collision of black holes consists of the bifurcation
of one black hole into two. 
This would be realized through an $(n-1)$-handle attachment, 
if such a process were possible (Fig.~\ref{wn-1-handle}).
It is, however, well known that
such a process is forbidden.~\cite{HE}
In general, the time reversal of the $\lambda$-handle attachment corresponds to
$(n-\lambda)$-handle attachment.

Before discussing general cases, let us consider the structure of a handlebody.
Recall that a $\lambda$-handle consists of the product space $D^\lambda\times D^{n-\lambda}$. 
The subset 
$D^\lambda\times \{0\}
\subset D^\lambda\times D^{n-\lambda}$
is called the core of the handlebody, 
and $\{0\}\times D^{n-\lambda}\subset D^\lambda\times D^{n-\lambda}$
is called the co-core.
The core and co-core intersect transversely at a point. 
This point can be regarded as a critical point $p$.

Let us refer to
the subset $W_s(p)$ of $\widetilde H$ 
\begin{equation}
W_s(p)=\{q\in M|\lim_{t\to +\infty}\exp_q tX=p\}
\end{equation}
which consists of points that converge to 
$p$ along the flow generated by the gradient-like vector field $X$,
as the stable manifold
with respect to the critical point $p$.
The stable manifold
$W_s(p)$ is homeomorphic to ${\mathbb R}^\lambda$ if 
the index of $p$ is given by $\lambda$.~\cite{Sm61}
Similarly, let us refer to the subset $W_u(p)\subset \widetilde H$ consisting of points
which converge to $p$ along the flow generated by $(-X)$ as the unstable manifold
with respect to $p$. 
For the unstable manifold, $W_u(p)\approx {\mathbb R}^{n-\lambda}$ holds.  
The portions of 
the stable and  unstable manifolds
in the handlebody can be
regarded as corresponding to the core and co-core, respectively.

The effect of smoothing the event horizon $H$ to $\widetilde H$ is to deform the null
vector field generating $H$ into a gradient-like vector field $X$.
The primary difference between the null geodesic generators and the flow generated by $X$
is that the former does not have future endpoints, but the latter can.
Thus, there are admissible and inadmissible processes for the smoothed manifold $\tilde H$.
An admissible process is given by $\widetilde H$, which is obtained from an 
in priciple realizable 
event horizon, while an inadmissible one is constructed from a spurious event horizon,
i.e., one that consists of the null hypersurface
containing null geodesic generators with a future endpoint.

\section{The structure of the critical points}\label{s3}

The spatial topology of a black hole changes only when the time function 
takes a critical value.
The time evolution of the black hole topology can be understood by considering its
local structure around critical points.
To determine whether a given topological change is admissible or inadmissible,
it is not sufficient to 
consider only the intrinsic structure of the event horizon.
Rather, it is required to take account of its imbedding structure relative 
to the space-time.

In a time slice, any point separate from $\widetilde H$ 
belongs to either of the black hole or the exterior of the black hole region.
It is useful to consider the local behavior of 
the black hole region or the exterior region near the critical point $p$.
Let us call the exterior ${\mathscr E}$ of the black hole region 
simply the exterior region, for brevity.
The exterior region is slightly deformed by the smoothing procedure.
The deformed exterior region is denoted by $\widetilde{{\mathscr E}}$, and the deformed
exterior region at the time $t$ by 
\begin{eqnarray}
\widetilde{{\mathscr E}}(t)=\widetilde{{\mathscr E}}\cap \Sigma (t)
=\Sigma (t)\setminus \overline{\widetilde{{\mathscr B}}(t)}.
\end{eqnarray}
The $0$-handle is placed at some $t\ge t(p)$.
Such an attachment describes the emergence of
the black hole region at the critical point $p$ and its expansion with time.
The emergence of a bubble, which consists of a part of $J^-({\mathscr I}^+)$, 
in the background of the black hole region would also be described by a $0$-handle
attachment.
This, however, never occurs, 
as we explain below in detail.
Hence, a $0$-handle attachment always describes the creation of a black hole
homeomorphic to the $n$-disk.

An $n$-handle attachment corresponds to the time reversal of a $0$-handle attachement.
This process, however, never occurs in real black hole space-time.
An $n$-handle is defined for $t\le t(p)$, which means that it terminates at the critical point $p$.
The crease set is isolated into critical points during the course of the smoothing procedure.
The gradient-like vector field, which can be regarded as being tangent 
to the generator of the deformed event horizon $\widetilde H$,
may have several inward (converging) directions 
at the critical point due to this smoothing procedure, 
while the original null generator of the
event horizon does not have an  inward direction at the crease set.
In the case of the $n$-handle, all the directions become 
inward
 at the critical point.
This implies that the null generators of the event horizon $H$
must have future endpoints at the critical point,
which is, of course, impossible.
It is thus seen that an $n$-handle attachment never occurs in  real 
black hole space-times.

The remaining cases are $\lambda$-handle attachments for $1\le \lambda\le n-1$.
In these cases, the $\lambda$-handle lies on either side of the critical point $p$ 
both in the future [$t>t(p)$] and in the past [$t<t(p)$].
Then, we consider the case in which 
the handle exists during the sufficiently small time interval 
$t\in [t(p)-\delta,t(p)+\delta]$ $(\delta>0)$,
 to understand the topological change of the black 
hole region at the critical point $p$.

\begin{figure}
\includegraphics[width=0.4\linewidth]{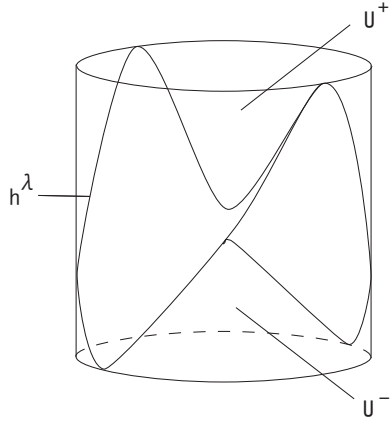}
\caption{The neighborhood $U$ of $p$ is separated by $h^\lambda$ into 
the future region, $U^+$, and the past region, $U^-$.}
\label{U}
\end{figure}

First, we introduce a coordinate system $\{t,x^i\}$ $(i=1,\cdots,n)$ 
in the neighborhood $U$ of $p$, 
where $t$ is a given function of time, and $\{x^i\}$ 
is the extension over $U$ of the cannonical coordinate appearing in the Morse Lemma 
such that 
each curve $(x^1,\cdots,x^n)=[{\rm const}]$ is timelike in $U$.
We assume that  $U$  is the solid cylinder 
given by $t\in [t(p)-\delta,t(p)+\delta]$,
$\sum (x^i)^2 \le\delta$.
In this coordinate system, the $\lambda$-handle $h^\lambda$ is given by 
the saddle surface
\begin{eqnarray*}
t=t(p)-(x^1)^2-\cdots-(x^\lambda)^2+(x^{\lambda+1})^2+\cdots+(x^n)^2
\end{eqnarray*}
in $U$, 
which is an acausal set if the constant $\delta$ is taken sufficiently small,
since $h^\lambda$ is tangent to the space-like hypersurface $t=t(p)$ at $p$.
Therefore, $h^\lambda$ separates $U$ into two open subsets, the future and past
regions
$U^+$ and $U^-$ of $U$,  where $U^+$ and $U^-$ are the subsets lying chronological 
future and past, respectively, of $h^\lambda$:
 $U^\pm =I^\pm(h^\lambda)\cap U$.
Explicitly, the future and past regions $U^\pm$ are the regions satisfying
\begin{eqnarray*}
t\gtrless t(p)-(x^1)^2-\cdots-(x^\lambda)^2+(x^{\lambda+1})^2+\cdots+(x^n)^2
\end{eqnarray*}
in $U$, respectively (Fig.~\ref{U}).

Because the $\lambda$-handle is a subset of the black hole boundary $\widetilde{H}$, 
one of $U^\pm$ is contained in the black hole region, $\widetilde{{\mathscr B}}$,
and the other in the exterior region, $\widetilde{{\mathscr E}}$.
However, the future region  $U^+$ of $U$ is always included in the black hole region, 
i.e. $U^+\subset \widetilde{{\mathscr B}}$, and hence 
we have $U^-\subset\widetilde{{\mathscr E}}$,
since the horizon is the boundary of the past set, $J^-({\mathscr I}^+)$.
Therefore, the black hole region $\widetilde{{\mathscr B}}(t(p)-\epsilon)\cap U$ in $U$
 at the time
$t=t(p)-\epsilon$ 
just before the critical time is
given by
\begin{eqnarray*}
(x^1)^2+\cdots+(x^\lambda)^2>(x^{\lambda+1})^2+\cdots+(x^n)^2+\epsilon,
\end{eqnarray*}
which is homotopic to the $(\lambda-1)$-sphere $S^{\lambda-1}$.
(For $\lambda=1$, $S^0$  simply consists of two points.)
Similarly, $\widetilde{{\mathscr B}}(t(p)+\epsilon)\cap U$
just after the critical time is given by
\begin{eqnarray*}
(x^1)^2+\cdots+(x^\lambda)^2+\epsilon>(x^{\lambda+1})^2+\cdots+(x^n)^2,
\end{eqnarray*}
which is homotopic to the $n$-disk.
In this way, the black hole region restricted to the small neighborhood of the critical point
$p$ is initially  homotopic to a sphere. Then, the internal region of the sphere
is filled up at the critical time $t=t(p)$ and eventually becomes
 homotopically trivial.
The exterior region, $\widetilde{{\mathscr E}}(t)\cap U$,
in $U$
is initially homotopic to an $n$-disk for $t=t(p)-\epsilon$.
Then, its $(n-\lambda)$-dimensional direction is penetrated by 
the black hole region at $t=t(p)$,
and thus it becomes homotopic to an $(n-\lambda-1)$-sphere $S^{n-\lambda-1}$ 
for $t=t(p)+\epsilon$.
If the spurious event horizon is also taken into account, 
the future region $U^+$ might be a subset of $\widetilde{{\mathscr E}}$, and therefore
the past region $U^-$ might be a subset of $\widetilde{{\mathscr B}}$.
Then, the black hole region in the $\lambda$-handle
might be homotopic to an $n$-disk initially and become homotopic to an $(n-\lambda-1)$-sphere 
finally,
and  vice versa for the exterior region.
Let us refer to such a topological change of the black hole region 
$\widetilde{{\mathscr B}}(t)\cap U$
from a region homotopic to a sphere to a region homotopic to a disk
as a black handle attachment, 
and that from a region homotopic to a disk to a region homotopic to the sphere 
as a white handle
attachment.
The above observation shows that only a black handle attachment
occurs if a sufficiently small neighborhood of the critical point is considered.
For example, a collision of black holes corresponds to a black $1$-handle attachment,
while the bifurcation of a black hole corresponds to a white 
$(n-1)$-handle
attachment in the sense that the homotopy type of the exterior region 
$\widetilde{{\mathscr E}}(t)\cap U$ changes
from that of $S^{n-2}$ to that of $D^n$. 
This local argument also elucidates te reason that 
a black hole collision is admissible while a black hole bifurcation, which is its
time reversal, is inadmissible.
We also note that the effect of  time reversal
is to convert a black $\lambda$-handle attachment into
a white $(n-\lambda)$-handle attachment.

It is appropriate to refer to the $0$-handle attachment
corresponding to the creation of a black hole
as a black $0$-handle attachment.
Then, the proposition above also applies to a $0$-handle attachment.

\section{Connectedness of the exterior region}\label{s4}
\begin{figure}[b]
\includegraphics[width=0.75\linewidth]{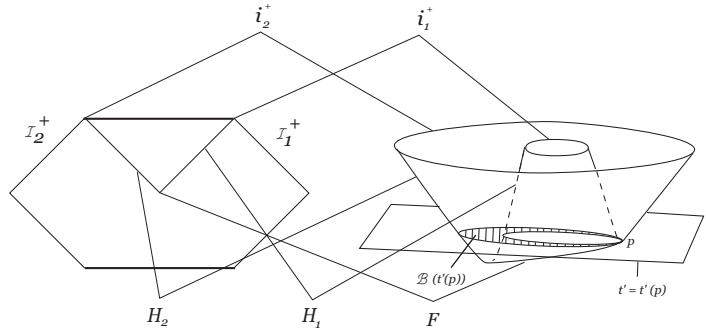}
\caption{The figure on the left is a conformal diagram of the maximally extended
Schwarzschild space-time. The structure of the event horizon defined with respect to the
two asymptotic ends is depicted on the right, with one dimension omitted. The shaded region
represents the black hole region at the critical time $t=t(p)$.
This corresponds to the 2-handle attachment, where the exterior region
is separated into a pair of connected components.}
\label{bifurcate}
\end{figure}

There also exist  processes
that are unrealizable due to global conditions.
Let us, for a moment, consider the event horizon 
in maximally extended Schwarzschild space-time.
Though we are interested in the event horizon defined
with respect to a specific asymptotic end, 
for the purpose of explanation, we examine the event horizon defined 
with respect to a pair of asymptotic ends in
Schwarzschild space-time (Fig.~\ref{bifurcate}).

Let ${\mathscr I}^+_1$ and ${\mathscr I}^+_2$  be the pair of future null infinities of
the maximally extended Schwarzschild space-time.
The event horizon here is defined by $H=\partial J^-({\mathscr I}^+_1\cup{\mathscr I}^+_2)$,
which is nondifferentiable at the bifurcate horizon $F=\partial J^-({\mathscr I}^+_1)
\cap \partial J^-({\mathscr I}^+_2)$.
Let $t$ be a global time function and $\chi$ be a global radial coordinate function such that
each two-surface $t$, $\chi=[{\rm const}]$ is invariant under the $SO(3)$ isometry.
These coordinates are chosen such that 
the bifurcation surface $F$ is located at $t=\chi=0$ and 
the event horizon $H$ is determined  by $t=|\chi|$ around $F$.
The smoothed event horizon $\widetilde H$ is also taken to be invariant under
the $SO(3)$ isometry.
Due to the symmetry of the configuration, the time function $t$ has critical points of
degenerate type. In fact, any point on bifurcate horizon $F$ is critical.
Here, we are not interested in such a nongeneric situation.
Instead, we consider a slightly different time slicing determined by the new time function
\begin{eqnarray*}
t'=t+\epsilon\sin^2{\vartheta\over 2},
\end{eqnarray*}
 where $\epsilon>0$ is a sufficiently small positive constant
and $\vartheta$, which satisfies $0\le \vartheta\le \pi$,
 is the usual polar coordinate of the 2-sphere.
Then, there appears only a pair of isolated critical
points at the north pole $(\vartheta=0$) and the south pole $(\vartheta=\pi)$
on the bifurcate horizon $F$, and the time function
$t'$ becomes the Morse function on $\widetilde H$.
At the time $t'=0$, the black hole appears at the north pole.
This is the $0$-handle attachment. The black hole formed there 
grows into a geometrically thick spherical shell with a hole at the south pole,
which is nevertheless a topological 3-disk.
At the time $t'=\epsilon$, the puncture at the south pole is filled, and
the black hole region becomes topologically
 $S^2\times [0,1]$.
The deformed event horizon $\widetilde H$ splits into a disjoint union of a pair of 2-spheres. 
This is the $2$-handle attachment.

This kind of $2$-handle attachment occurs because the event horizon is defined with
respect to the two asymptotic ends, which is in general inadmissible 
if the future null infinity is connected, as we assume from this point.
To understand the above statement, it should be noted that there is no process
through which the several connected components 
of the exterior  region $\widetilde{{\mathscr E}}(t)=\widetilde{{\mathscr E}}\cap \Sigma(t)$
at time $t$
merge together at a later time because such a handle attachment 
is not admissible.
It is also seen that no connected component of $\widetilde{{\mathscr E}}(t)$
 disappears, because possible $n$-handle
attachments are inadmissible. These facts imply that the number of 
connected components of the exterior region $\widetilde{{\mathscr E}}(t)$
cannot decrease with the time function $t$.
On the other hand, there is only one connected component of 
the exterior region  $\widetilde{{\mathscr E}}(t)$ for sufficiently large $t$,
because of the connectedness of ${\mathscr I}^+$.
This observation shows that 
the exterior region $\widetilde{{\mathscr E}}(t)$ remains connected in any process.

The only possible process through which the number of  connected components of the
exterior region $\widetilde{{\mathscr E}}(t)$ changes is an $(n-1)$-handle attachment,
as constructed above in the Schwarzschild space-time.
This is because the subset
$D^\lambda\times\partial D^{n-\lambda}$
of the boundary of the $\lambda$-handle
\begin{eqnarray*}
\partial h^\lambda\approx (\partial D^\lambda\times D^{n-\lambda})\cup
(D^\lambda\times\partial D^{n-\lambda}),
\end{eqnarray*}
namely the part of $\partial h^\lambda$ which is the complement 
of the preimage of the attaching map
\begin{eqnarray*}
f: \partial h^{\lambda}\supset \partial D^\lambda\times D^{n-\lambda}
\to \widetilde{H}_t,
\end{eqnarray*}
is disconnected only when $\lambda=n-1$.
In this case, the homotopy type of the exterior region $\widetilde{{\mathscr E}}(t)$
changes from that of 
an $n$-disk to that of $S^0$, namely  two points.
Note, however, that this does not imply that 
the exterior region $\widetilde{{\mathscr E}}(t)$ is always separated into two 
disconnected parts
through the $(n-1)$-handle attachment.
For example, a transition from 
the black ring horizon $\approx S^{n-2}\times S^1$ to the spherical black hole horizon
$\approx S^{n-1}$ is realized through a black $(n-1)$-handle attachment, which 
 pinches the longitude $\{\mbox{a point}\}\times S^1\subset S^{n-2}\times S^1$
into a point.
The exterior region $\widetilde{{\mathscr E}}(t)$ remains connected all the while.
Thus, there are both admissible and inadmissible processes for  $(n-1)$-handle attachments.
An $(n-1)$-handle attachment is inadmissible
if it separates 
the exterior region $\widetilde{{\mathscr E}}(t)$.

\section{Concluding remarks}\label{s5}
The arguments given in this paper are summerized by the following rules.
Assume that 
(i)
\begin{em}
an $(n+1)$-dimensional space-time $M$ is asymptotically flat
and the future null infinity ${\mathscr I}^+$ is connected, or
the event horizon $H=\partial J^-({\mathscr I}^+)$ is 
defined with respect to a single asymptotic end,
\end{em}
(ii) 
\begin{em}
the space-time $M$ admits a smooth global time function $t$,
\end{em}
(iii) 
\begin{em}
the event horizon $H$ can be
deformed 
so that the  black hole $\widetilde{{\mathscr B}}(t)$ deformed accordingly
at each time $t$ is smooth and homeomorphic to
original one ${\mathscr B} (t)$ at each time $t$
and the time function $t$ becomes the Morse function
on $\widetilde H$.
\end{em}
Then, the topological evolution of the event horizon can be regarded as a $\lambda$-handle
attachment $(0\le\lambda\le n)$ subject to the following rules:
\begin{enumerate}
\begin{em}
\item The $n$-handle attachment is inadmissible.
\item Only the black $\lambda$-handle attachment $(0\le \lambda\le n-1)$,
where the black hole region in the neighborhood of the critical point
varies from  the region homotopic to the sphere $S^{\lambda-1}$ (regarded as 
the empty set for $\lambda=0$) to the $n$-disk $D^n$,
is admissible.
\item The $(n-1)$-handle attachment which separates the spatial section of the
exterior region of the black hole is inadmissible.
\end{em}
\end{enumerate}

The first rule simply states that no connected component of a black hole disappears.
It also implies that if a bubble of the exterior region forms within the black hole region,
it does not vanish.

The second rule is concerned with the imbedding structure of the event horizon
relative to the space-time manifold.
The  neighborhood of the critical point 
is separated into two regions by the event horizon. 
One changes homotopically from a sphere to a disk and the other from a disk to a sphere.
We call it a black handle attachment when the former corresponds to the black hole region
and a white handle attachment otherwise.
Then, the second rule  states that a white handle attachment never occurs.
The reverse process, in which a black hole region homotopically changes from
a disk to a sphere, is ruled out.
A white $0$-handle attachment, which describes the emergence of the exterior region,
is also forbidden.
This gives another reason for the well-known result that
a black hole cannot bifurcate, because it corresponds to a white
$(n-1)$-handle attachment.

\begin{figure}
\includegraphics[width=0.6\linewidth]{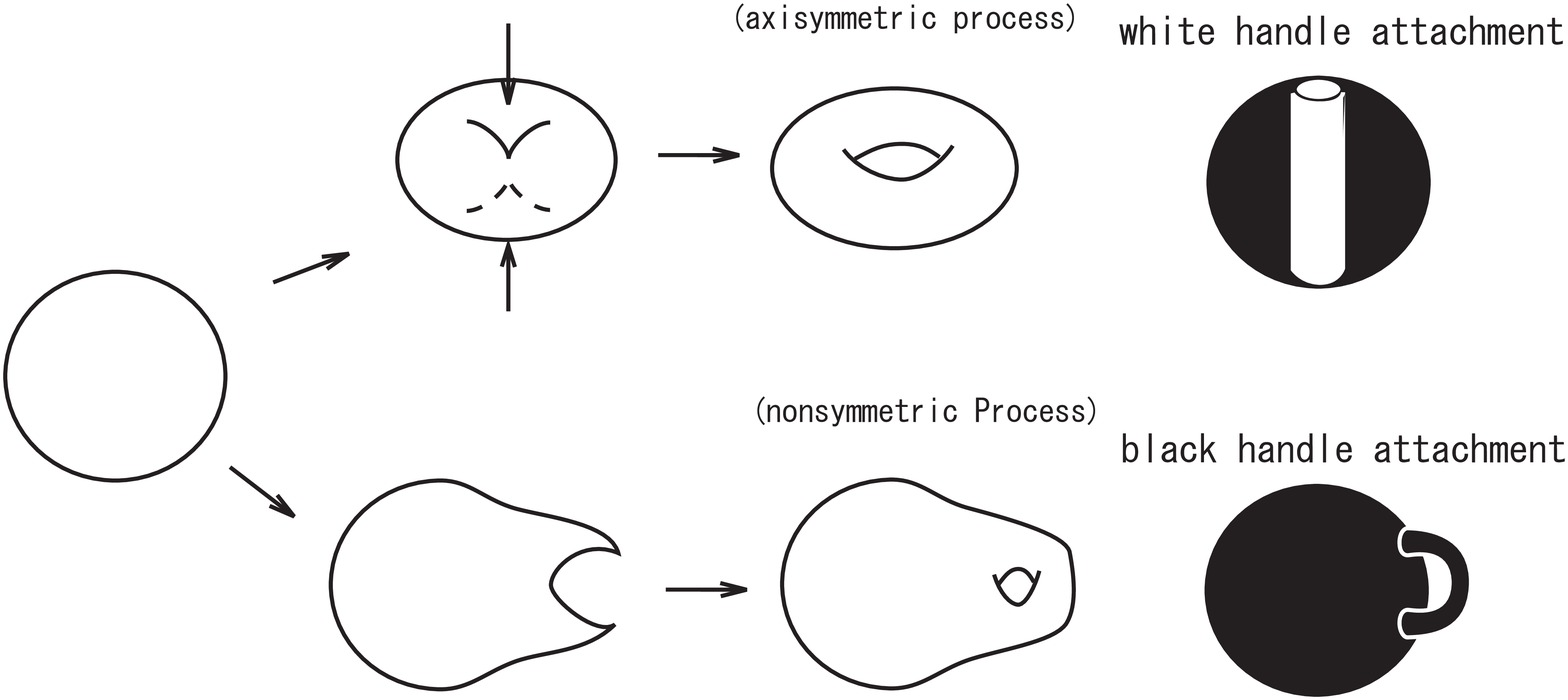}
\caption{Black ring formation from a spherical black hole must be non-axisymmetric in 
 real black hole space-times.}
\label{ring_formation}
\end{figure}

The second rule applies to more general situations.
For example, let us consider the topological evolution of the event horizon from
$S^{n-1}$ to $S^{n-2}\times S^1$ in $(n+1)$-dimensional space-time $(n\ge 3)$.
When it is realized with a single critical point, it corresponds to a $1$-handle attachment.
Here, one might expect two possibilities if the second rule is not considered.
One possibility is that the $1$-handle is attached in the exterior region of the black hole.
This is locally equivalent to the merging of a pair of black holes, where
these two black holes are connected elsewhere irrelevant.
The other possibility is that it is attached from the inside
such that the $1$-handle pierces the black hole region.
In asymptotically flat space-times, only the latter includes axisymmetric configurations
such that a spherical black hole is pinched out along the symmetric axis;
here the axisymmetric configuration is such that 
the space-time possesses the $SO(n-1)$ isometry and 
the time slicing respects this symmetry.
However, this latter possibility corresponds to a white $1$-handle attachment,
which is impossible, and only the former, which corresponds to a black $1$-handle attachment,
is possible.
In particular, a transition from a spherical event horizon $(\approx S^{n-1})$
to a black ring horizon $(\approx S^{n-2}\times S^1)$ in asymptotically flat space-times
is always non-axisymmetric
in the sense that such a configuration cannot possess $SO(n-1)$ symmetry
(Fig.~\ref{ring_formation}).

While the apparent horizon must be diffeomorphic to a two-sphere
in four-dimensional space-times
under the dominant energy condition,
a torus event horizon may appear, even under the dominant energy condition,
via a black $1$-handle attachment to the spherical horizon.
More generally, an event horizon with an arbitrary number of genura may be formed by
several black $1$-handle attachments.

The third rule is not directly determined by the local structure of the critical point.
It states that the exterior region ${\mathscr E}(t)={\mathscr E}\cap \Sigma(t)$
at each time is always connected under the
assumption that ${\mathscr I}^+$ is connected.
Thus, the possibility that there forms a bubble of the exterior region inside the
 black hole horizon is ruled out. It  should, however, be noted that such a process is
possible if ${\mathscr I}^+$ consists of several connected components.
This may also be related to the topological censorship theorem.~\cite{Friedman:1993ty}
The topological censorship theorem states that all causal curves from ${\mathscr I}^-$
to ${\mathscr I}^+$ are homotopic under the null energy condition.
This also forbids the formation of a bubble of the exterior region inside the black hole,
because otherwise there would be two nonhomotopic causal curves from ${\mathscr I}^-$
to ${\mathscr I}^+$, one passing inside the horizon and the other  outside.
Our argument, however, does not depend on energy conditions.


\end{document}